\documentclass[pra,preprint,longbibliography]{revtex4-1}
\usepackage{graphicx}	
\usepackage{appendix}
\usepackage{braket}
\usepackage{rotating}
\usepackage{simplewick}
\usepackage{amssymb}
\usepackage{amsmath}
\usepackage{txfonts}
\usepackage{bm}
\usepackage{color}
\usepackage{multirow}
\usepackage{booktabs}
\usepackage{dcolumn}

\begin{document}

\title{Converging Finite-Temperature Many-Body Perturbation Theory In The Grand Canonical Ensemble That Conserves The Average Number Of Electrons} 

\author{So Hirata}\email{sohirata@illinois.edu.}
\author{Punit K. Jha}

\affiliation{ 
Department of Chemistry, University of Illinois at Urbana-Champaign, Urbana, Illinois 61801, USA}

\date{\today}

\begin{abstract}
A finite-temperature perturbation theory for the grand canonical ensemble is introduced that expands the 
chemical potential in a perturbation series and conserves the average number of electrons, ensuring
 charge neutrality of the system at each perturbation order. Two (sum-over-state and reduced) classes of 
analytical formulas are obtained in a straightforward, algebraic, time-independent derivation 
for the first-order corrections to the chemical potential, grand potential, and internal energy, with the aid of several identities of the Boltzmann sums also introduced in this study.
These formulas are numerically verified against benchmark data from thermal full configuration interaction.
For a nondegenerate ground state, the finite-temperature perturbation theory reduces analytically to and is consistent with 
the M{\o}ller--Plesset perturbation theory as temperature ($T$) tends to zero. For a degenerate ground state,
it should instead reduce to the Hirschfelder--Certain degenerate perturbation theory  as $T \to 0$.
\end{abstract}

\maketitle 

\section{Introduction} 

In the previous Chapter \cite{JhaHirata}, we showed that the finite-temperature perturbation theory for electrons described 
in a number of modern textbooks \cite{thouless1972quantum,mattuck1992guide,march1995many,Fetter} and excellent reviews \cite{bloch,balian,blochbook,SANTRA}
does not reproduce the benchmark data obtained as the $\lambda$-derivatives of the exact
finite-temperature theory \cite{Kou} with the perturbation-scaled Hamiltonian, $\hat{H} = \hat{H}_0 + \lambda \hat{V}$.
The discrepancy was traced \cite{JhaHirata,WhiteChan} not so much to mathematical issues but to a misuse 
of the ensemble. The theory given in textbooks (beyond the zeroth-order Fermi--Dirac theory) adopts a grand canonical ensemble that 
holds chemical potential fixed at some arbitrary value and allows the average number of electrons to fluctuate,
violating charge neutrality as a basic tenet of equilibrium thermodynamics \cite{Levin}. 
Note that the energy of a system with long-range unscreened interactions such as a charged plasma is not 
thermodynamically extensive \cite{Fisher,Dyson,HirataARPC}.
As a consequence, the perturbation theory described in textbooks does not 
converge at the exact limit of the neutral system. Worse still,  beyond the zeroth order, 
it describes a massively charged system to which equilibrium thermodynamics is no longer applicable.

We need a finite-temperature perturbation theory that also expands the chemical
potential in a perturbation series so as to ensure charge neutrality at any perturbation order and thereby 
converges at the exact, thermal full-configuration-interaction (FCI) \cite{Kou} limits. 
In this Chapter, we present such a perturbation theory and derive analytical formulas of its first-order corrections to the chemical
potential, grand potential, and internal energy. Two classes of analytical formulas 
were derived in an easy-to-follow, time-independent, algebraic (nondiagrammatic) manner:\ The sum-over-states formulas 
 are correct in all cases, while the more compact, reduced formulas are valid
  only for a nondegenerate ground state at any temperature or for 
a degenerate or nondegenerate ground state at a nonzero temperature.
We show that both formulas reproduce the benchmark data obtained as the $\lambda$-derivatives of the thermal FCI results in the case of a nondegenerate ground state.

There have been a persistent question as to whether the zero-temperature limits of a finite-temperature perturbation theory for 
internal energy should agree with the M{\o}ller--Plesset (MP) perturbation theory \cite{shavitt} for energy at respective orders. 
Kohn and Luttinger  \cite{kohn} were the first to pose a related (if not the same) question, taking the finite-temperature second-order perturbation 
theory for a homogeneous electron gas (having a degenerate ground state) as an example. We show that, for a nondegenerate ground state, the finite-temperature perturbation theory introduced here
is consistent with the MP perturbation theory in the sense that the first-order internal-energy correction reduces to the first-order MP energy correction both  analytically and numerically.
For a degenerate ground state, however, the finite-temperature perturbation theory is {\it not} consistent with the MP perturbation theory; it instead reduces to and is consistent with the 
Hirschfelder--Certain (HC) {\it degenerate} perturbation theory \cite{hirschfelder}.
 
\section{Finite-temperature perturbation theory in grand canonical ensemble} 

The electronic grand partition function \cite{Kou} per molecule of an ideal gas of identical molecules at given temperature $T$ is 
\begin{eqnarray}
\Xi = \sum_{I=1}^{2^m} e^{-\beta E_I  + \beta \mu {N}_I}, \label{Xi}
\end{eqnarray}
where $\beta = (k_{\text{B}}T)^{-1}$, $E_I$ is the FCI energy of the $I$th state and $N_I$ is the number of electrons in the same state. 
The summation over a capital letter index ($I$) runs over all (exponentially many) $2^{m}$ states 
spanned by $m$ spinorbital basis functions. Chemical potential 
$\mu$ is determined by solving
\begin{eqnarray}
\bar{N} = \frac{\sum_{I=1}^{2^m} N_I e^{-\beta E_I + \beta \mu {N}_I}}{\sum_{I=1}^{2^m} e^{-\beta E_I  + \beta \mu {N}_I}}, \label{Nbar}
\end{eqnarray}
where $\bar{N}$ is a given average number of electrons that keeps the system electrically neutral. 
These two are the equations of state to be solved simultaneously for $\Xi$ and $\mu$.
A valid grand canonical ensemble for electrons 
holds $\bar{N}$ fixed at the value that maintains neutrality of the system by adjusting $\mu$ (not vice versa). This is not a choice
but a nonnegotiable requirement, without which the system is massively charged, causing the energy per particle to be divergent \cite{Fisher,Dyson,HirataARPC} 
and equilibrium thermodynamics to break down \cite{Levin}.

Grand potential $\Omega$ and internal energy $U$ per molecule are related to $\Xi$ by
\begin{eqnarray}
\Omega &=& -\frac{1}{\beta} \ln \Xi, \\
U &=& -\frac{\partial}{\partial \beta} \ln \Xi + \mu \bar{N}, 
\end{eqnarray}
and also to each other by
\begin{eqnarray}
\Omega &=& U - T S - \mu \bar{N}, \label{OmegaU}
\end{eqnarray}
where $S$ is entropy.

We expand $\Xi$ and $\mu$ as well as $\Omega$ and $U$ in converging perturbation series,
\begin{eqnarray}
\Xi &=& \Xi^{(0)} + \lambda \Xi^{(1)} + \lambda^2 \Xi^{(2)}+ \lambda^3 \Xi^{(3)} + \dots, \\
\mu &=& \mu^{(0)} + \lambda \mu^{(1)} + \lambda^2 \mu^{(2)}+ \lambda^3 \mu^{(3)} + \dots, 
\end{eqnarray}
and so forth, and seek to write each correction in terms of the perturbation corrections of the energy,
\begin{eqnarray}
E_I &=& E_I^{(0)} + \lambda E_I^{(1)} + \lambda^2 E_I^{(2)}+ \lambda^3 E_I^{(3)} + \dots,  \label{expandE}
\end{eqnarray}
where $E_I^{(n)}$ is the $n$th-order correction to the $I$th-state energy according to the 
HC degenerate perturbation theory \cite{hirschfelder}, of which the MP perturbation
theory \cite{shavitt} for a nondegenerate ground state is a special case.

They conform to the canonical definition of a perturbation theory, in which the $n$th-order
correction of quantity $X$ is  
\begin{eqnarray}
X^{(n)} &=& \left. \frac{1}{n!} \frac{\partial^n X(\lambda)}{\partial \lambda^n}\right|_{\lambda=0},  \label{lambda}
\end{eqnarray}
where $X(\lambda)$ is the value of $X$ (such as $E_I$, $\Xi$, $\mu$, $\Omega$, and $U$) 
calculated by FCI with the Hamiltonian $\hat{H}_0 + \lambda\hat{V}$. 
We adopt the M{\o}ller--Plesset partitioning and, therefore, the zeroth-order Hamiltonian, $\hat{H}_0$, is the sum of the (zero-temperature) Fock operators plus the nuclear repulsion energy.
We emphasize again that the perturbation corrections to energies that exactly match this canonical definition 
are furnished, in the most general case, by the HC degenerate perturbation theory and not by the MP perturbation theory
(because many excited, ionized, and electron-attached states are exactly degenerate at the zeroth order 
even if the ground state is not).

We have obtained numerically these corrections at lower orders and shown that
the finite-temperature perturbation theory described in textbooks does not reproduce such benchmark data \cite{JhaHirata} because
it neglects to expand $\mu$ in a perturbation series, thereby allowing $\bar{N}$ to fluctuate and causing
the system to be massively charged. We call the method that evaluates these derivatives 
by finite-difference formulas the $\lambda$-variation method \cite{Hirata2017,JhaHirata}.
In this study, we will perform the differentiation 
analytically to arrive at the corresponding analytical formulas for $\mu^{(1)}$, $\Omega^{(1)}$, and $U^{(1)}$.

In the following, we use the Taylor series,
\begin{eqnarray}
e^{a+b} &=& e^a + e^ab  +  e^a\frac{b^2}{2!} +  e^a\frac{b^3}{3!} + \dots, \label{exp} \\
\ln(a+b) &=& \ln a + \frac{b}{a} - \frac{b^2}{2a^2} + \frac{b^3}{3a^3} + \dots, \label{log}
\end{eqnarray}
which are rapidly convergent when $a \gg b$. We will also use four identities of the Boltzmann sums, which will be introduced as needed.

\section{Zeroth order} 

The zeroth-order theory \cite{Fetter} in the grand canonical ensemble is correct
in the sense that chemical potential $\mu^{(0)}$ is adjusted to keep the average number of electrons constant 
at $\bar{N}$. We review a time-independent and algebraic (as opposed to time-dependent or diagrammatic) derivation of this theory
as a preparation for its application to the first-order theory.

\subsection{Fermi--Dirac theory}

The zeroth-order grand partition function is given by
\begin{eqnarray}
\Xi^{(0)}  &=& \sum_{I=1}^{2^m} e^{-\beta E_I^{(0)} + \beta \mu^{(0)} N_I} = \sum_{I=1}^{2^m} \xi^{(0)}_I \label{Xi00}
\end{eqnarray}
with
\begin{eqnarray}
\xi^{(0)}_I = e^{-\beta E_I^{(0)} + \beta \mu^{(0)} N_I}
\end{eqnarray}
and 
\begin{eqnarray}
E_I^{(0)} = E_{\text{nuc.}} + \sum_i^{\text{occ.}} \epsilon_i^{(0)};~~~N_I = \sum_i^{\text{occ.}} 1,
\end{eqnarray}
where $E_{\text{nuc.}}$ is the nuclear repulsion energy
and $\epsilon_i^{(0)}$ is the $i$th canonical Hartree--Fock (HF) spinorbital energy.
The summation over small letter index ($i$) runs over spinorbitals occupied in the $I$th Slater determinant,
which is a solution of the zeroth-order Schr{\"o}dinger equation. The occupied orbitals should not be confused with
those occupied in the HF ground state; they vary depending on state $I$.

This can be simplified to
\begin{eqnarray}
\Xi^{(0)}  &=& e^{-\beta E_{\text{nuc.}}} \prod_i^{\text{all}} \sum_{n=0}^{1} (e^{-\beta \epsilon_i^{(0)} + \beta \mu^{(0)}})^n \\
&=&e^{-\beta E_{\text{nuc.}}} \prod_i^{\text{all}} \left\{1 + e^{\beta(\mu^{(0)}-\epsilon_i^{(0)})}\right\}, \label{Xi0}
\end{eqnarray}
where products over $i$ run over all $m$ spinorbitals spanned by the basis set.
Equation (\ref{Nbar}) also reduces to
\begin{eqnarray}
\bar{N} &=& \frac{\sum_{I=1}^{2^m} N_I e^{-\beta E_I^{(0)} + \beta \mu^{(0)} N_I}}
{\sum_{I=1}^{2^m} e^{-\beta E_n^{(0)} + \beta \mu^{(0)} N_I}} \label{22} \\
&=& \frac{1}{\beta} \frac{\partial}{\partial \mu^{(0)}} \ln \sum_{I=1}^{2^m} e^{-\beta E_I^{(0)} + \beta \mu^{(0)} N_I} \\
&=& \frac{1}{\beta} \frac{\partial}{\partial \mu^{(0)}} \ln 
e^{-\beta E_{\text{nuc.}}} \prod_i^{\text{all}} \left\{1 + e^{\beta(\mu^{(0)}-\epsilon_i^{(0)})}\right\} 
\\
&=&  \sum_i^{\text{all}} \frac{e^{\beta(\mu^{(0)}-\epsilon_i^{(0)})}}{1 + e^{\beta(\mu^{(0)}-\epsilon_i^{(0)})}} 
= \sum_i^{\text{all}} f_i^- , \label{Nbar0}
\end{eqnarray}
where $f_i^-$ is the Fermi--Dirac occupancy function of the $i$th spinorbital,
\begin{eqnarray}
 f_i^- =   \frac{1}{1 + e^{\beta(\epsilon_i^{(0)}-\mu^{(0)})}} . \label{FD}
\end{eqnarray}

Using $\Xi^{(0)}$ of equation  (\ref{Xi0}) with $\mu^{(0)}$ determined by solving equation (\ref{Nbar0}), we obtain the zeroth-order
grand potential and internal energy as 
\begin{eqnarray}
\Omega^{(0)} &=& -\frac{1}{\beta} \ln \Xi^{(0)} =E_{\text{nuc.}} +  \frac{1}{\beta}\sum_i^{\text{all}} \ln f_i^+, \\
U^{(0)} &=& -\frac{\partial}{\partial \beta} \ln \Xi^{(0)} + \mu^{(0)} \bar{N} = E_{\text{nuc.}} + \sum_i^{\text{all}} \epsilon_i^{(0)} f_i^-,
\end{eqnarray}
where 
$f_i^+$ is the Fermi--Dirac vacancy function of the $i$th spinorbital,
\begin{eqnarray}
 f_i^+ =  1 - f_i^- . \label{FDV}
\end{eqnarray}

\subsection{Boltzmann-sum identity I}

It is instructive to rederive equation (\ref{Nbar0}) in a more general manner. Let $a_I = \sum_i ^{\text{occ.}} a_i$.
Using a notation, $\nu_i = \beta(\mu^{(0)}-\epsilon_i^{(0)})$, and canceling the common factor of $e^{-\beta E_{\text{nuc.}}}$
in the numerator and denominator, we can prove an identity,
{\scriptsize
\begin{eqnarray}
\frac{\sum_{I=1}^{2^m} a_I e^{-\beta E_I^{(0)} + \beta \mu^{(0)} N_I}}
{\sum_{I=1}^{2^m} e^{-\beta E_I^{(0)} + \beta \mu^{(0)} N_I}} 
&=&
\frac{\sum_{i=1}^m a_i e^{\nu_i} + \sum_{i < j}^m (a_i+a_j) e^{\nu_i}e^{\nu_j} + \sum_{i<j<k}^m (a_i + a_j + a_k) e^{\nu_i}e^{\nu_j}e^{\nu_k} + \dots
+ \sum_{i_1 < \dots < i_m}^m (a_{i_1} + \dots + a_{i_m}) e^{\nu_{i_1}}\cdots e^{\nu_{i_m}} }
{(1+e^{\nu_1})(1+e^{\nu_2})\cdots(1+e^{\nu_m})}
\nonumber\\
&=& \sum_{i=1}^m a_i f_i^-;~(\text{Identity I}), \label{boltzmann1}
\end{eqnarray}
}
by mathematical induction.
Equation (\ref{Nbar0}) is nothing but Identity I with $a_i = 1$ for any $i$. 

\section{First order} 

\subsection{Sum-over-states analytical formulas}

An application of equation (\ref{exp}) to equation (\ref{Xi}) leads to the following expression for the first-order correction to grand partition function $\Xi$:
\begin{eqnarray}
\Xi^{(1)}  &=&  \sum_{I=1}^{2^m} e^{-\beta E_I^{(0)} + \beta \mu^{(0)} N_I} (-\beta E_I^{(1)} + \beta \mu^{(1)} N_I)
= \sum_{I=1}^{2^m}  \xi^{(1)}_{I}  \label{Xi1}
\end{eqnarray}
with
\begin{eqnarray}
\xi^{(1)}_{I} = \xi^{(0)}_I (-\beta E_{I}^{(1)} + \beta \mu^{(1)} N_{I}),
\end{eqnarray}
where $E_I^{(1)}$ is the first-order correction to the energy of the $I$th state according to the HC degenerate perturbation theory \cite{hirschfelder}.

The same procedure to equation (\ref{Nbar}) leads to the equation for $\mu^{(1)}$ as
\begin{eqnarray}
\bar{N} &=& \frac{ \sum_{I=1}^{2^m} N_I \xi_I^{(1)} }
{\sum_{I=1}^{2^m} \xi_I^{(1)}  },  \label{N1}
\end{eqnarray}
which can be solved for $\mu^{(1)}$ as
\begin{eqnarray}
\mu^{(1)} &=& \frac{ \sum_{I=1}^{2^m} \xi_I^{(0)} E_I^{(1)} (N_I- \bar{N})}
{\sum_{I=1}^{2^m} \xi_I^{(0)} N_I (N_I- \bar{N})}.  \label{N2}
\end{eqnarray}

Using equation (\ref{log}), we obtain the first-order corrections to the grand potential and internal energy as
\begin{eqnarray}
\Omega^{(1)} &=& -\frac{1}{\beta}\frac{\Xi^{(1)}}{\Xi^{(0)}} =- \frac{1}{\beta} \frac{\sum_{I=1}^{2^m} \xi_I^{(1)} }{\sum_{I=1}^{2^m} \xi^{(0)}_I}, \label{Omega1} \\
U^{(1)} &=& -\frac{\partial}{\partial \beta} \frac{\Xi^{(1)}}{\Xi^{(0)}} + \mu^{(1)}\bar{N} \label{U1} \\ 
&=& \Omega^{(1)} + \mu^{(1)}\bar{N} + \beta \Omega^{(1)} (U^{(0)}  - \mu^{(0)}\bar{N}) \nonumber\\
&& - \frac{\beta}{\Xi^{(0)}} \sum_{I=1}^{2^m} \xi^{(0)}_I ( E_I^{(0)}  - \mu^{(0)}N_I)( E_I^{(1)}  -\mu^{(1)}N_I).  \label{USOS}
\end{eqnarray}
In the last equality, we used
\begin{eqnarray}
\frac{\sum_{I=1}^{2^m} \xi^{(0)}_I (E_I^{(0)}  -  \mu^{(0)}N_I) }{\Xi^{(0)}}  = U^{(0)}  - \mu^{(0)}\bar{N},
\end{eqnarray}
which can be readily proven with Identity I [equation (\ref{boltzmann1})]. 

The third and fourth terms of equation (\ref{USOS}) contain products of extensive quantities such as $\Omega^{(1)} U^{(0)}$, which are individually 
non-size-consistent \cite{HirataTCA}. However, it can be shown (see below) that they mutually cancel exactly across the two terms, leaving
only rigorously extensive contributions; the first-order finite-temperature perturbation theory is size-consistent in spite of these terms.

The aforementioned equations are the correct and complete (if not compact) 
analytical formulas of the first-order finite-temperature perturbation
theory in the grand canonical ensemble applicable to either a degenerate or nondegenerate ground state at any temperature down to zero.
Equation (\ref{N2}) is first solved for $\mu^{(1)}$ using $\{E_I^{(1)}\}$ 
unambiguously defined and calculable by the HC degenerate perturbation theory \cite{hirschfelder}.
Once $\mu^{(1)}$ is determined, $\Xi^{(1)}$ is evaluated with equation (\ref{Xi1}). 
Next, $\Omega^{(1)}$ and $U^{(1)}$ are computed with equations (\ref{Omega1}) and (\ref{USOS}), respectively, using
$\mu^{(1)}$ and $\{E_I^{(1)}\}$ as well as the zeroth-order quantities. We have verified these formulas numerically (see the Appendix).

\subsection{Boltzmann-sum identities II, III, and IV}

Here, we introduce three more identities for the Boltzmann sums, which are useful for deriving compact reduced formulas. The second identity reads
\begin{widetext}
\begin{eqnarray}
&& 
\frac{\sum_{i=1}^m a_i b_i e^{\nu_i} + \sum_{i < j}^m (a_i+a_j)(b_i+b_j) e^{\nu_i}e^{\nu_j} + \dots
+ \sum_{i_1 < \dots < i_m}^m (a_{i_1} + \dots + a_{i_m}) (b_{i_1} + \dots + b_{i_m}) e^{\nu_{i_1}}\cdots e^{\nu_{i_m}} }
{(1+e^{\nu_1})(1+e^{\nu_2})\cdots(1+e^{\nu_m})}
\nonumber\\&& 
= \sum_{i=1}^m a_i b_i f_i^- f_i^+ + \sum_{i=1}^m  a_i f_i^- \sum_{i=1}^m b_i  f_i^-;~(\text{Identity II}). \label{boltzmann2}
\end{eqnarray}
The third identity is 
\begin{eqnarray}
&&
\frac{\sum_{i < j}^mc_{ij} e^{\nu_i}e^{\nu_j} + \sum_{i<j<k}^m (c_{ij}+c_{ik}+c_{jk}) e^{\nu_i}e^{\nu_j}e^{\nu_k} + \dots
+ \sum_{i_1 < \dots < i_m}^m (c_{i_1i_2} + \dots + c_{i_{m-1}i_m}) e^{\nu_{i_1}}\cdots e^{\nu_{i_m}} }
{(1+e^{\nu_1})(1+e^{\nu_2})\cdots(1+e^{\nu_m})}
\nonumber\\&& 
= \sum_{i<j}^m c_{ij} f_i^-f_j^-;~(\text{Identity III}), \label{boltzmann3}
\end{eqnarray}
where $c_{ij} = c_{ji}$ for any $i$ and $j$. Note that $c_{ii}$ is absent in any summation. 
Identities II and III can be proven by mathematical induction with Identity I. 
Finally, the fourth reads
{\scriptsize
\begin{eqnarray}
&&
\frac{\sum_{i < j}^m (b_i + b_j) c_{ij} e^{\nu_i}e^{\nu_j} + \sum_{i<j<k}^m (b_i+b_j+b_k)(c_{ij}+c_{ik}+c_{jk}) e^{\nu_i}e^{\nu_j}e^{\nu_k} + \dots
+ \sum_{i_1 < \dots < i_m}^m (b_{i_1} + \dots + b_{i_m}) (c_{i_1i_2} + \dots + c_{i_{m-1}i_m}) e^{\nu_{i_1}}\cdots e^{\nu_{i_m}} }
{(1+e^{\nu_1})(1+e^{\nu_2})\cdots(1+e^{\nu_m})}
\nonumber\\&& 
= \sum_{i<j}^m (b_i f_i^+ + b_j f_j^+ ) c_{ij} f_i^-f_j^- + \sum_{i=1}^m b_i f_i^-  \sum_{i<j}^m c_{ij} f_i^-f_j^- ;~(\text{Identity IV}), \label{boltzmann4}
\end{eqnarray}}
\end{widetext}
which can also be proven by mathematical induction with the aid of Identities I, II, and III. These proofs are elementary and not reproduced here.

\subsection{Reduced analytical formulas}

The foregoing analytical formulas for the first-order perturbation theory involve sum over exponentially many states, and are inconvenient to use. 
For a system with a nondegenerate ground state or at a nonzero temperature, we can reduce them into much more compact formulas involving 
only molecular integrals in the spirit of the Fermi--Dirac theory or (incorrect) diagrammatic finite-temperature perturbation theory given in 
textbooks. 

First, we rewrite equation (\ref{Omega1}) as
\begin{eqnarray}
\Omega^{(1)} &=& 
\frac{\sum_{I=1}^{2^m}  e^{-\beta E_I^{(0)} + \beta \mu^{(0)} N_I} ( - \beta E_I^{(1)} + \beta \mu^{(1)} N_I )}
{\sum_{I=1}^{2^m}  e^{E_I^{(0)} - \mu^{(0)} N_I}}. \label{Xi0red0}
\end{eqnarray}
Here, $E_I^{(1)}$ is the first-order HC perturbation correction to the energy \cite{hirschfelder}
and is written as
\begin{eqnarray}
E_I^{(1)} = \langle \Psi_I^{(0)} | \hat{V} | \Psi_I^{(0)} \rangle,
\end{eqnarray}
where $\Psi_I^{(0)}$ is the zeroth-order wave function and $\hat{V}$ is the perturbation operator in the M{\o}ller--Plesset partitioning.

For nondegenerate $I$th state, $\Psi_I^{(0)}$ is a single Slater determinant $\Phi_I$, making the evaluation of the above integral trivial:
\begin{eqnarray}
E_I^{(1)} = \langle \Phi_I | \hat{V} | \Phi_I \rangle =  \sum_{i < j}^{\text{occ.}} \langle ij || ij \rangle
- \sum_{i}^{\text{occ.}} \langle i||i \rangle, \label{V}
\end{eqnarray}
with 
\begin{eqnarray}
\langle i||i \rangle = \sum_{j=1}^{\text{HOMO}} \langle ij||ij \rangle,
\end{eqnarray}
where `occ.'\ again refers to the spinorbitals occupied in the $I$th Slater determinant, whereas  
`HOMO' stands for the highest-occupied spinorbital of the $\bar{N}$-electron ground state (hereafter simply the ground state); the 
last summation runs over spinorbitals occupied in the ground state. Each four-index integral is a usual 
antisymmetrized two-electron integral \cite{shavitt}.

When the $I$th state is degenerate (in the sense that there are $d$ states sharing the same zeroth-order energy), 
$\Psi_I^{(0)}$ is no longer a Slater determinant, but 
a linear combination of $d$ degenerate Slater determinants, $\{ \Phi_J \}$ \cite{hirschfelder}.
\begin{eqnarray}
\Psi_I^{(0)} = \sum_{J}^{d} \Phi_J U_{JI},
\end{eqnarray}
where $U_{JI}$ is an element of a unitary matrix, which can only be determined by a highly involved procedure of the HC degenerate perturbation theory \cite{hirschfelder}.
It is more a rule than an exception that the zeroth-order energies of Slater determinants display degeneracy. 

However, thanks to the orthonormality of $\{\Psi_I^{(0)}\}$ within each degenerate group, the following equality holds:
\begin{eqnarray}
\frac{1}{d} \sum_{J}^{d} E_J^{(1)} =  \sum_{i < j}^{\text{occ.}} \langle ij || ij \rangle
- \sum_{i}^{\text{occ.}}  \langle i||i \rangle , \label{sumd}
\end{eqnarray}
where the summation over $J$ is taken for all $d$ degenerate states. In other words, although $E_J^{(1)}$ for the 
$J$th individual state cannot be written simply as the right-hand side of the above equation, its average 
over all $d$ degenerate states can be \cite{SANTRA}.

Therefore, if all $E_J^{(1)}$ from a degenerate group are summed with an equal weight, 
one obtains the correct final formula by pretending as if equation (\ref{V}) were true for each state.
This is indeed the case with equation (\ref{Xi0red0}) if the ground state is nondegenerate or temperature is nonzero. 
For a degenerate ground state at zero temperature, the states that are degenerate with the ground state are summed 
with an unequal weight and the use of equation (\ref{sumd}) becomes inappropriate. We will consider this case in Sec.\ \ref{firstdeg}, but
in this section we confine ourselves to a nondegenerate ground state or nonzero temperature.

We can then write equation (\ref{V}) as
\begin{eqnarray}
E_I^{(1)} - \mu^{(1)} N_I = \sum_{i}^{\text{occ.}} a_i + \sum_{i < j}^{\text{occ.}} c_{ij} 
\end{eqnarray}
with
\begin{eqnarray}
a_i = -  \langle i||i \rangle - \mu^{(1)};~~~c_{ij} =\langle ij || ij \rangle, \label{ca}
\end{eqnarray}
and $c_{ii} = 0$. Using Identities I and III [equation (\ref{boltzmann1}) and (\ref{boltzmann3})],
we can reduce equation (\ref{Xi0red0}) into
\begin{eqnarray}
\Omega^{(1)} &=& \frac{1}{2} \sum_{i,j}^{\text{all}} \langle ij || ij \rangle f_i^-f_j^-
- \sum_{i}^{\text{all}} \left(  \langle i||i \rangle +  \mu^{(1)} \right) f_i^-  \label{Omegafinal2}  \\
&=& \frac{1}{2} \sum_{i,j}^{\text{all}} \langle ij || ij \rangle f_i^-f_j^-
- \sum_{i}^{\text{all}} \langle i||i \rangle  f_i^- -   \mu^{(1)} \bar{N}, \label{Omegafinal}
\end{eqnarray}
where the summations run over all spinorbitals.

With the same choice of $\{a_i\}$ and $\{c_{ij}\}$ in equation (\ref{ca}) as well as
$b_i = 1$ for any $i$, equation (\ref{N1}) for $\mu^{(1)}$ is simplified with Identities II and IV 
[equation (\ref{boltzmann2}) and (\ref{boltzmann4})] as follows: 
\begin{widetext}
\begin{eqnarray}
\bar{N} = \frac{\frac{1}{2} \sum_{i,j}^{\text{all}} \langle ij || ij \rangle f_i^-f_j^- (\bar{N} + f_i^+ + f_j^+)
- \sum_{i}^{\text{all}} \langle i||i \rangle f_i^- (\bar{N} + f_i^+)
- \sum_{i}^{\text{all}} \mu^{(1)} f_i^- (\bar{N} + f_i^+)}
{\frac{1}{2} \sum_{i,j}^{\text{all}} \langle ij || ij \rangle f_i^-f_j^-
- \sum_{i}^{\text{all}} \langle i||i \rangle f_i^- 
- \sum_{i}^{\text{all}} \mu^{(1)} f_i^-}, \label{Nbarreduced}
\end{eqnarray}
which can be solved for $\mu^{(1)}$ as
\begin{eqnarray}
\mu^{(1)} = 
\frac{\frac{1}{2} \sum_{i,j}^{\text{all}} \langle ij || ij \rangle f_i^-f_j^- ( f_i^+ + f_j^+)
- \sum_{i}^{\text{all}}  \langle i||i \rangle f_i^- f_i^+}
{\sum_{i}^{\text{all}}  f_i^-f_i^+}. \label{Nbarreduced2}
\end{eqnarray}
\end{widetext}
The last equation can also be obtained by a direct application of Identities I--IV to equation (\ref{N2}).

Likewise, applying Identities I--IV to the sum-over-states formula for $U^{(1)}$ [equation (\ref{USOS})], 
we first observe exact mutual cancellation of non-size-consistent terms and then arrive at
a compact analytical formula, 
\begin{eqnarray}
U^{(1)} &=& \frac{1}{2} \sum_{i,j}^{\text{all}} \langle ij || ij \rangle f_i^-f_j^-
- \sum_{i}^{\text{all}}  \langle i||i \rangle f_i^- \nonumber \\
&& + \frac{\beta}{2} \sum_{i,j}^{\text{all}}  \langle ij || ij \rangle f_i^-f_j^-
\left\{ (\mu^{(0)}- \epsilon_i^{(0)} ) f_i^+ + (\mu^{(0)}- \epsilon_j^{(0)} )f_j^+\right\} \nonumber\\
&&- \beta \sum_{i}^{\text{all}}   \left( \langle i||i \rangle + \mu^{(1)} \right) (\mu^{(0)}- \epsilon_i^{(0)} ) f_i^- f_i^+ , \label{Ufinal} 
\end{eqnarray}
which no longer contains a non-size-consistent term that is a product of two or more extensive quantities. 
The same expression can be obtained alternatively \cite{SANTRA} by returning to equation (\ref{U1}), namely, 
\begin{eqnarray}
U^{(1)} &=&- \frac{\partial}{\partial \beta} \left( - \beta\Omega^{(1)} \right) + \mu^{(1)}\bar{N}  \\
&=&\Omega^{(1)} + \beta \frac{\partial \Omega^{(1)}}{\partial \beta} + \mu^{(1)}\bar{N}, 
\end{eqnarray}
and using 
\begin{eqnarray}
\frac{\partial f_i^- }{\partial \beta} 
= (\mu^{(0)}- \epsilon_i^{(0)} ) f_i^- f_i^+ . 
\end{eqnarray}
Care must be exercised that partial differentiation with $\beta$ acts only on $f_i^-$, $f_i^+$, and $\bar{N}$ via equation (\ref{Omegafinal2}), 
but not on $\mu^{(1)}$.

It should be noted that the foregoing reduced formulas are valid only for a nondegenerate ground state or at a nonzero temperature, 
whereas the sum-over-states formulas and underlying HC degenerate perturbation theory should be correct in all cases.

The corresponding reduced formulas \cite{SANTRA} for $\Omega^{(1)}$ and $U^{(1)}$ according to the textbook ansatz wherein
$\mu$ is held fixed at $\mu^{(0)}$ are special cases of the more general expressions reported here and are 
readily obtained by substituting $\mu^{(1)}=0$ in equations (\ref{Omegafinal}) and (\ref{Ufinal}),
respectively. 

\section{Comments on related methods}

The finite-temperature HF \cite{thouless1972quantum,mattuck1992guide,march1995many,Fetter}, finite-temperature 
Green's function \cite{Zgid1,Zgid2}, and thermal coupled-cluster \cite{sanyal,mandal2,mandal,WhiteChan2} theories
as well as thermal FCI theory \cite{Kou} all have a provision to adjust $\mu^{(0)}$ so as to restore the correct average number of electrons, making the calculation physically sound
and their results more in line with the correct values \cite{JhaHirata}.

It is then also possible to adjust $\mu^{(0)}$ by some root-finding procedure in the framework of
the textbook finite-temperature perturbation theory to restore the net charge neutrality \cite{SANTRA}.
Then, the theory forms a converging series of approximations towards thermal FCI. However, this series may not be considered a canonical perturbation series 
defined by equation (\ref{lambda}) because adjusting $\mu^{(0)}$ amounts to altering the partitioning of the Hamiltonian at each perturbation order, 
obscuring the order of perturbation corrections (in the same sense that HF theory involving rotation of orbitals 
cannot be equated to the first-order perturbation theory). In other words, this series differs from the series presented in this work that gives closed expressions for the first-order corrections to all thermodynamic quantities of interest, which furthermore agree exactly with the $\lambda$-variation results.

\section{Zero-temperature limit and degeneracy\label{firstdeg}}

\subsection{Zeroth order}

In the zero-temperature ($\beta \to \infty$) limit, we can analytically show \cite{Kou} and numerically confirm \cite{JhaHirata}
\begin{eqnarray}
\lim_{\beta \to \infty} \mu^{(0)} = \frac{\epsilon^{(0)}_{\text{HOMO}} + \epsilon^{(0)}_{\text{LUMO}}}{2},
\end{eqnarray}
where HOMO again stands for the highest-occupied spinorbital of the $\bar{N}$-electron ground state
and LUMO the lowest-unoccupied spinorbital. 

In this limit, the Fermi--Dirac occupancy function is
\begin{eqnarray}
f_i^- = 1 ~ (i \leq \text{HOMO}); ~~~f_i^- = 0 ~ (i > \text{HOMO}), \label{zeroTf}
\end{eqnarray}
assuming that the spinorbitals are in an ascending order of energy. 
However, a special consideration \cite{PedersonJackson} may become necessary when the ground state is degenerate. 
In this case, 
\begin{eqnarray}
\lim_{\beta \to \infty} \mu^{(0)} = \epsilon^{(0)}_{\text{HOMO}} = \epsilon^{(0)}_{\text{LUMO}},
\end{eqnarray}
leaving $f_{\text{HOMO}}^-$ and $f_{\text{LUMO}}^-$ (as well as the Fermi--Dirac functions of all other degenerate spinorbitalss)
undefined, although their sum must take a certain value 
that maintains the average number of electrons at $\bar{N}$, say,
\begin{eqnarray}
f_{\text{HOMO}}^- + f_{\text{LUMO}}^- = 1,
\end{eqnarray}
in the case of doubly degenerate HOMOs sharing one electron on average. 
However, a partitioning between $f_{\text{HOMO}}^-$ and $f_{\text{LUMO}}^-$ is indeterminate. 

Nevertheless, we can unambiguously determine the zero-temperature limits of 
$\Omega^{(0)}$ and $U^{(0)}$ as
\begin{eqnarray}
\lim_{\beta \to \infty} \Omega^{(0)} &=& \sum_{i=1}^{\text{HOMO}} \epsilon_i^{(0)} - \mu^{(0)}\bar{N}, \\
\lim_{\beta \to \infty} U^{(0)} &=& \sum_{i=1}^{\text{HOMO}} \epsilon_i^{(0)}, \label{limU0}
\end{eqnarray}
because they do not depend on the partitioning of the Fermi--Dirac occupancy functions among the degenerate HOMOs.
These are consistent with the zeroth-order MP perturbation theory and also with each other according to equation (\ref{OmegaU}).

\subsection{First order}

Next, we consider the zero-temperature limits of the first-order corrections. We divide the problem into two cases:\ a nondegenerate ground state 
versus a degenerate ground state.

In a nondegenerate ground state, the reduced analytical formulas such as equations (\ref{Nbarreduced}), (\ref{Omegafinal}), and (\ref{Ufinal}) are valid at any temperature down to zero. 
Using equation (\ref{zeroTf}) and the identity \cite{kohn,hirata_KL,SANTRA},
\begin{eqnarray}
\lim_{\beta \to \infty} \beta f_i^- f_i^+  = \delta\left(\epsilon^{(0)}_i - \mu^{(0)}\right),
\end{eqnarray}
we find
\begin{eqnarray}
\lim_{\beta \to \infty} \mu^{(1)} &=& 0, \\
\lim_{\beta \to \infty} \Omega^{(1)} &=& - \frac{1}{2} \sum_{i,j=1}^{\text{HOMO}} \langle ij || ij \rangle \\
\lim_{\beta \to \infty} U^{(1)} &=& - \frac{1}{2} \sum_{i,j=1}^{\text{HOMO}} \langle ij || ij \rangle \equiv E_{\text{MP1}} \label{MP1} ,
\end{eqnarray}
the last of which is the familiar first-order MP perturbation energy correction at zero temperature, $E_{\text{MP1}}$. 
These are numerically verified in the Appendix.
For a nondegenerate ground state, therefore, there is no inconsistency between the finite-temperature perturbation theory and MP perturbation theory.

When the ground state is $d$-fold degenerate ($d > 1$), at zero temperature, they are no longer summed at an equal Boltzmann weight, and
the reduced analytical formulas derived with equation (\ref{sumd}) become invalid. We, therefore, have to revert to the sum-over-states analytical
formulas. However, at first glance, even the sum-over-states formulas such as equation (\ref{Xi1}) appear to suffer from the same issue of ambiguous 
Boltzmann weights; at zero temperature, $\beta = \infty$ and $E_1^{(0)} - \mu^{(0)}N_1 = 0$, rendering the exponent $-\beta E_1^{(0)} + \beta \mu^{(0)}N_1$ indeterminate. 

The correct zero-temperature behavior of the exponents can be inferred by considering the $\lambda$-dependence of the thermal FCI energies
or the closely related HC degenerate perturbation corrections to the energies of degenerate states. States that are degenerate at the zeroth order or $\lambda=0$ 
become nondegenerate or have a lesser degree of degeneracy as $\lambda$ is raised from zero in thermal FCI. The perturbation corrections
to the internal energy at $T=0$ are, by definition, the $\lambda$-derivatives of the lowest energy (not the average energy of the degenerate ground states at $\lambda=0$). Correspondingly, the HC degenerate perturbation corrections to energies
are generally different from one another for degenerate states, gradually lifting degeneracy as the perturbation order is raised \cite{hirschfelder}. Then, for the first-order correction 
to be consistent with the $\lambda$-derivative of the corresponding thermal FCI result, the sum over states should be interpreted as being dominated by
the zeroth-order state whose first-order energy correction is the most negative.  
Applying this logic to $U^{(1)}$ [equation (\ref{USOS})], we predict
\begin{eqnarray}
\lim_{\beta \to \infty} U^{(1)} &=& E_{\text{HC1}},
\end{eqnarray}
where $E_{\text{HC1}}$ is the most negative among the first-order HC degenerate perturbation corrections to the degenerate zeroth-order ground states.

This equation indicates that there is no inconsistency between the finite-temperature perturbation theory and HC degenerate perturbation theory,
but there is an inconsistency (reminiscent of the Kohn--Luttinger inconsistency \cite{kohn,hirata_KL,SANTRA})
between the finite-temperature perturbation theory and MP perturbation theory because 
$E_{\text{HC1}}$ is not equal to $E_{\text{MP1}}$. The former, $E_{\text{HC1}}$, is determined 
by the complex, but unambiguous procedure \cite{hirschfelder} outlined by Hirschfelder and Certain and 
cannot be expressed simply as the right-hand side of 
equation (\ref{MP1}). 

The compact formulas for $U^{(1)}$ [equation (\ref{Ufinal})] and even for $\Omega^{(1)}$ [equation (\ref{Omegafinal})] appear to reduce analytically
to the MP1 expression at zero temperature in all cases including a degenerate ground state. We consider such an interpretation   
incorrect for three reasons:\ (1) these formulas are derived using an assumption [equation (\ref{sumd})] that is 
not valid for a degenerate ground state at zero temperature, (2) $\mu^{(1)}$ is not necessarily zero in this case, 
and (3) the correct zero-temperature limit
is not the MP perturbation theory but the HC degenerate perturbation theory for a degenerate ground state.

We should note that the foregoing argument in the case of degenerate ground states
has not been numerically verified unlike the conclusion for nondegenerate ground states, which is numerically confirmed
in the Appendix. It is based on an assumption of the availability of strictly degenerate reference wave functions,
which are not uniquely determined by the zeroth-order theory. However, only the rate of convergence of the HC degenerate perturbation series 
and thus of the finite-temperature many-body perturbation theory varies depending on the reference, while
the converged limits should be the same correct thermal FCI limits.

\section{Summary}

Original contributions of this work are summarized as follows:

(1) We have introduced a converging finite-temperature perturbation theory in the grand canonical ensemble 
that expands chemical potential in a perturbation series and 
conserves the average number of electrons, thereby ensuring charge neutrality of the system at each perturbation order.

(2) We have obtained the sum-over-states analytical formulas for the first-order corrections of the chemical potential, grand potential, and 
internal energy, which are valid for a degenerate or nondegenerate ground state and at any temperature down to zero. 
They are given as equations (\ref{N2}), (\ref{Omega1}), and (\ref{USOS}), respectively.
These formulas depend crucially on the HC degenerate perturbation theory, whose details
are described elsewhere \cite{hirschfelder}.

(3) We have reported the compact, reduced analytical formulas for the same quantities 
as equations (\ref{Nbarreduced2}), (\ref{Omegafinal}), and (\ref{Ufinal}), respectively,
which are valid for a nondegenerate ground state or at a nonzero temperature. 
We have established a straightforward, time-independent, nondiagrammatic derivation strategy of these formulas
by introducing several identities for Boltzmann sums in the grand canonical ensemble.

(4) We shed new light on the issue of the zero-temperature limit of the converging finite-temperature perturbation theory. 
The finite-temperature perturbation theory introduced here is consistent with the MP perturbation theory for
a nondegenerate ground state. For a degenerate ground state, 
the finite-temperature theory is consistent with the HC degenerate perturbation theory, but not with the MP perturbation theory.
For a degenerate ground state, the compact analytical formulas such as equation (\ref{Ufinal}) are correct at $T > 0$, but 
suddenly become incorrect at $T=0$; the sum-over-states
formulas should be used instead.

\section{Future outlook}
This theory should be extended to the second and higher orders in the grand canonical ensemble as well as 
to the canonical ensemble \cite{JhaHirata2019}.
These results should be generalized to a recursive algebraic definition in the style of the Rayleigh--Sch\"{o}dinger 
perturbation theory, if possible, which should permit a general-order implementation and 
establish and justify diagrammatic rules for the grand potential and internal energy as well as their size-consistency 
through a time-indepdendent linked-diagram theorem \cite{shavitt,Hirata2017}.

\acknowledgments
This work was supported by the Center for Scalable, Predictive methods for Excitation and Correlated phenomena (SPEC), which is funded by the U.S. Department of Energy, Office of Science, Office of Basic Energy Sciences, Chemical Sciences, Geosciences, and Biosciences Division, as a part of the Computational Chemical Sciences Program and
also by the U.S. Department of Energy, Office of Science, Office of Basic Energy Sciences under Grant No.\ DE-SC0006028.

\appendix*
\section{Numerical verification}

\begin{table*}
\caption{ \label{tab:only} Comparison of the first-order corrections to chemical potential ($\mu^{(1)}$), 
grand potential ($\Omega^{(1)}$), and internal energy ($U^{(1)}$) obtained with the $\lambda$-variation, sum-over-states formulas,
and reduced analytical formulas  
as a function of temperature ($T$) for the hydrogen fluoride molecule (0.9168~\AA) in the STO-3G basis set.}
{\scriptsize
\begin{ruledtabular}
\begin{tabular}{lddddddddd}
&\multicolumn{3}{c}{$\mu^{(1)} / E_{\text{h}}$}& \multicolumn{3}{c}{$\Omega^{(1)} / E_{\text{h}}$} & \multicolumn{3}{c}{$U^{(1)} / E_{\text{h}}$} \\ \cline{2-4} \cline{5-7} \cline{8-10}
&\multicolumn{1}{c}{$\lambda$-variation\footnotemark[1]}  
&\multicolumn{1}{c}{SoS\footnotemark[2]}  
&\multicolumn{1}{c}{Analytical} 
&\multicolumn{1}{c}{$\lambda$-variation\footnotemark[1]}  
&\multicolumn{1}{c}{SoS\footnotemark[2]}  
&\multicolumn{1}{c}{Analytical} 
&\multicolumn{1}{c}{$\lambda$-variation\footnotemark[1]}  
&\multicolumn{1}{c}{SoS\footnotemark[2]}  
&\multicolumn{1}{c}{Analytical}  \\
{$T /~\text{K}$} 
&\multicolumn{1}{c}{Eq.~(\ref{lambda})}  
&\multicolumn{1}{c}{Eq.~(\ref{N2})}  
&\multicolumn{1}{c}{Eq.~(\ref{Nbarreduced2})} 
&\multicolumn{1}{c}{Eq.~(\ref{lambda})}  
&\multicolumn{1}{c}{Eq.~(\ref{Omega1})}  
&\multicolumn{1}{c}{Eq.~(\ref{Omegafinal})} 
&\multicolumn{1}{c}{Eq.~(\ref{lambda})}  
&\multicolumn{1}{c}{Eq.~(\ref{USOS})}  
&\multicolumn{1}{c}{Eq.~(\ref{Ufinal})}  \\ \hline
$10^4$ & 0.0000 & 0.0000 & -0.0000 & -45.9959 & -45.9961\footnotemark[3] & -45.9959 & -45.9959 & -45.9959 & -45.9959  \\
$10^5$ & -0.0752 & -0.0752 & -0.0752 & -45.2684 & -45.2684 & -45.2684 & -45.9479 & -45.9479 & -45.9479 \\
$10^6$ & -0.1690 & -0.1690 & -0.1690 & -44.5256 & -44.5256 & -44.5256 & -46.1767 & -46.1767 & -46.1767 \\
$10^7$ & -0.2981 & -0.2981 & -0.2981 & -43.1991 & -43.1991 & -43.1991 & -46.2355 & -46.2355 & -46.2355 \\
$10^8$ & -0.4122 & -0.4122 & -0.4122 & -41.9847 & -41.9847 & -41.9847 & -46.1180 & -46.1180 & -46.1180 \\
\end{tabular}
\footnotetext[1]{Obtained using a central seven-point finite difference. See also Jha and Hirata \cite{JhaHirata}.}
\footnotetext[2]{Obtained using the sum-over-states formulas with $\{E_I^{(1)}\}$ computed 
by the $\lambda$-variation method using forward seven- through nine-point finite differences.}
\footnotetext[3]{This value seems to suffer from a slightly greater numerical error.}
\end{ruledtabular}
}
\end{table*}

Table \ref{tab:only} documents the first-order corrections to $\mu$, $\Omega$, and $U$ in a converging perturbation series that expands 
$\mu$ perturbatively and maintains the average number of electrons at $\bar{N}$ at any perturbation order or at any value of perturbation strength ($\lambda$). 
The $\lambda$-variation data were already reported in the previous Chapter \cite{JhaHirata}.

At a lower temperature ($10^4$~K), a determination of $\mu^{(0)}$ by a bisection method becomes technically difficult, causing 
some numerical noise. Otherwise, the three methods are mutually in exact agreement, verifying the correctness of formulas and computational methods.
The results numerically confirm the zero-temperature limits at $T=10^4$~K. The perturbation theory described in
textbooks that holds $\mu$ fixed (e.g., at $\mu^{(0)}$) 
gives completely different values except at the low-temperature limit. 

%
\end{document}